\documentclass[amsmath,amssymb,superscriptaddress,floatfix,reprint,prd,nofootinbib]{revtex4-1}
\usepackage{graphicx}
\usepackage{bm}
\usepackage[]{units}
\usepackage{dcolumn}
\usepackage{multirow}
\usepackage{gensymb}
\usepackage[caption=false]{subfig}

\newcommand{\ssr}{Space Science Reviews}
\newcommand{\apjl}{ApJ Letters}
\newcommand{\jcap}{Journal of Cosmology and Astroparticle Physics}
\newcommand{\mnras}{Monthly Notices of the RAS} 

\begin{document}

\title{Atmospheric and Astrophysical Neutrinos above 1 TeV Interacting in IceCube}

\begin{abstract}

The IceCube Neutrino Observatory was designed primarily to search for
high-energy (TeV--PeV) neutrinos produced in distant astrophysical objects. A
search for $\gtrsim 100$~TeV neutrinos interacting inside the instrumented volume
has recently provided evidence for an isotropic flux of such neutrinos. At lower
energies, IceCube collects large numbers of neutrinos from the weak
decays of mesons in cosmic-ray air showers. Here we present the results of a
search for neutrino interactions inside IceCube's instrumented volume between
1~TeV and 1~PeV in 641 days of data taken from 2010--2012, lowering the energy
threshold for neutrinos from the southern sky below 10 TeV for the first time,
far below the threshold of the previous high-energy analysis. Astrophysical
neutrinos remain the dominant component in the southern sky down to a deposited energy of 10 TeV.
From these data we derive new constraints on the diffuse astrophysical neutrino
spectrum, $\Phi_{\nu} = 2.06^{+0.4}_{-0.3} \times 10^{-18}
\left({E_{\nu}}/{10^5 \,\, \rm{GeV}} \right)^{-2.46 \pm 0.12} {\rm {GeV^{-1} \,
cm^{-2} \, sr^{-1} \, s^{-1}} } $ for $25 \,\, \text{TeV} < E_{\nu} < 1.4 \,\, \text{PeV}$,
as well as the strongest upper limit yet on
the flux of neutrinos from charmed-meson decay in the atmosphere, 1.52 times
the benchmark theoretical prediction used in previous IceCube results at 90\%
confidence.

\end{abstract}


\affiliation{III. Physikalisches Institut, RWTH Aachen University, D-52056 Aachen, Germany}
\affiliation{School of Chemistry \& Physics, University of Adelaide, Adelaide SA, 5005 Australia}
\affiliation{Dept.~of Physics and Astronomy, University of Alaska Anchorage, 3211 Providence Dr., Anchorage, AK 99508, USA}
\affiliation{CTSPS, Clark-Atlanta University, Atlanta, GA 30314, USA}
\affiliation{School of Physics and Center for Relativistic Astrophysics, Georgia Institute of Technology, Atlanta, GA 30332, USA}
\affiliation{Dept.~of Physics, Southern University, Baton Rouge, LA 70813, USA}
\affiliation{Dept.~of Physics, University of California, Berkeley, CA 94720, USA}
\affiliation{Lawrence Berkeley National Laboratory, Berkeley, CA 94720, USA}
\affiliation{Institut f\"ur Physik, Humboldt-Universit\"at zu Berlin, D-12489 Berlin, Germany}
\affiliation{Fakult\"at f\"ur Physik \& Astronomie, Ruhr-Universit\"at Bochum, D-44780 Bochum, Germany}
\affiliation{Physikalisches Institut, Universit\"at Bonn, Nussallee 12, D-53115 Bonn, Germany}
\affiliation{Universit\'e Libre de Bruxelles, Science Faculty CP230, B-1050 Brussels, Belgium}
\affiliation{Vrije Universiteit Brussel, Dienst ELEM, B-1050 Brussels, Belgium}
\affiliation{Dept.~of Physics, Chiba University, Chiba 263-8522, Japan}
\affiliation{Dept.~of Physics and Astronomy, University of Canterbury, Private Bag 4800, Christchurch, New Zealand}
\affiliation{Dept.~of Physics, University of Maryland, College Park, MD 20742, USA}
\affiliation{Dept.~of Physics and Center for Cosmology and Astro-Particle Physics, Ohio State University, Columbus, OH 43210, USA}
\affiliation{Dept.~of Astronomy, Ohio State University, Columbus, OH 43210, USA}
\affiliation{Niels Bohr Institute, University of Copenhagen, DK-2100 Copenhagen, Denmark}
\affiliation{Dept.~of Physics, TU Dortmund University, D-44221 Dortmund, Germany}
\affiliation{Dept.~of Physics, University of Alberta, Edmonton, Alberta, Canada T6G 2E1}
\affiliation{Erlangen Centre for Astroparticle Physics, Friedrich-Alexander-Universit\"at Erlangen-N\"urnberg, D-91058 Erlangen, Germany}
\affiliation{D\'epartement de physique nucl\'eaire et corpusculaire, Universit\'e de Gen\`eve, CH-1211 Gen\`eve, Switzerland}
\affiliation{Dept.~of Physics and Astronomy, University of Gent, B-9000 Gent, Belgium}
\affiliation{Dept.~of Physics and Astronomy, University of California, Irvine, CA 92697, USA}
\affiliation{Dept.~of Physics and Astronomy, University of Kansas, Lawrence, KS 66045, USA}
\affiliation{Dept.~of Astronomy, University of Wisconsin, Madison, WI 53706, USA}
\affiliation{Dept.~of Physics and Wisconsin IceCube Particle Astrophysics Center, University of Wisconsin, Madison, WI 53706, USA}
\affiliation{Institute of Physics, University of Mainz, Staudinger Weg 7, D-55099 Mainz, Germany}
\affiliation{Dept.~of Physics and Astronomy, Michigan State University, East Lansing, MI 48824, USA}
\affiliation{Universit\'e de Mons, 7000 Mons, Belgium}
\affiliation{Technische Universit\"at M\"unchen, D-85748 Garching, Germany}
\affiliation{Bartol Research Institute and Dept.~of Physics and Astronomy, University of Delaware, Newark, DE 19716, USA}
\affiliation{Dept.~of Physics, University of Oxford, 1 Keble Road, Oxford OX1 3NP, UK}
\affiliation{Dept.~of Physics, Drexel University, 3141 Chestnut Street, Philadelphia, PA 19104, USA}
\affiliation{Physics Department, South Dakota School of Mines and Technology, Rapid City, SD 57701, USA}
\affiliation{Dept.~of Physics, University of Wisconsin, River Falls, WI 54022, USA}
\affiliation{Oskar Klein Centre and Dept.~of Physics, Stockholm University, SE-10691 Stockholm, Sweden}
\affiliation{Dept.~of Physics and Astronomy, Stony Brook University, Stony Brook, NY 11794-3800, USA}
\affiliation{Dept.~of Physics, Sungkyunkwan University, Suwon 440-746, Korea}
\affiliation{Dept.~of Physics, University of Toronto, Toronto, Ontario, Canada, M5S 1A7}
\affiliation{Dept.~of Physics and Astronomy, University of Alabama, Tuscaloosa, AL 35487, USA}
\affiliation{Dept.~of Astronomy and Astrophysics, Pennsylvania State University, University Park, PA 16802, USA}
\affiliation{Dept.~of Physics, Pennsylvania State University, University Park, PA 16802, USA}
\affiliation{Dept.~of Physics and Astronomy, Uppsala University, Box 516, S-75120 Uppsala, Sweden}
\affiliation{Dept.~of Physics, University of Wuppertal, D-42119 Wuppertal, Germany}
\affiliation{Dept.~of Physics, Yale University, New Haven, CT 06520, USA}
\affiliation{DESY, D-15735 Zeuthen, Germany}

\author{M.~G.~Aartsen}
\affiliation{School of Chemistry \& Physics, University of Adelaide, Adelaide SA, 5005 Australia}
\author{M.~Ackermann}
\affiliation{DESY, D-15735 Zeuthen, Germany}
\author{J.~Adams}
\affiliation{Dept.~of Physics and Astronomy, University of Canterbury, Private Bag 4800, Christchurch, New Zealand}
\author{J.~A.~Aguilar}
\affiliation{D\'epartement de physique nucl\'eaire et corpusculaire, Universit\'e de Gen\`eve, CH-1211 Gen\`eve, Switzerland}
\author{M.~Ahlers}
\affiliation{Dept.~of Physics and Wisconsin IceCube Particle Astrophysics Center, University of Wisconsin, Madison, WI 53706, USA}
\author{M.~Ahrens}
\affiliation{Oskar Klein Centre and Dept.~of Physics, Stockholm University, SE-10691 Stockholm, Sweden}
\author{D.~Altmann}
\affiliation{Erlangen Centre for Astroparticle Physics, Friedrich-Alexander-Universit\"at Erlangen-N\"urnberg, D-91058 Erlangen, Germany}
\author{T.~Anderson}
\affiliation{Dept.~of Physics, Pennsylvania State University, University Park, PA 16802, USA}
\author{C.~Arguelles}
\affiliation{Dept.~of Physics and Wisconsin IceCube Particle Astrophysics Center, University of Wisconsin, Madison, WI 53706, USA}
\author{T.~C.~Arlen}
\affiliation{Dept.~of Physics, Pennsylvania State University, University Park, PA 16802, USA}
% ...
\maketitle

\section{Introduction} 
\label{sec:introduction}

High-energy neutrinos are ideal cosmic messengers, produced whenever
cosmic rays interact with matter or photons near their as-yet unknown acceleration sites,
and carrying information about the conditions there to Earth without being
deflected by magnetic fields or absorbed by intervening matter
\cite{1995PhR...258..173G, LearnedAnnualReview, 2002RPPh...65.1025H,
Becker2008173}. At the same time,
neutrinos produced in cosmic-ray air showers provide information about hadronic
physics in kinematic regions that are difficult to probe with terrestrial
accelerators. Here we present an analysis of the diffuse flux of neutrinos
observed from 2010--2012 via events with vertices contained in the IceCube
detector \cite{DAQPaper} and depositing more than 1~TeV of energy. We derive
new constraints on the energy spectrum of the previously-observed \cite{HESE,
HESE_3year} astrophysical neutrino flux as well as the maximum contribution
from the decay of charmed mesons in the atmosphere covering significantly lower
energies. This will allow further limits on the possible scenarios
\cite{PhysRevD.88.043009, 2013JCAP...06..030C, 2013PhRvD..88h3007W, 2013PhRvL.111d1103K,
2013JCAP...01..028R, 2013arXiv1305.7404S, 2013PhRvD..87f3011H,
2013arXiv1306.5021A, 2013arXiv1310.7194G, 2013PhRvD..88h1302R}
that have been put forward to explain the observed astrophysical flux.

In the first section, we review the sources of neutrinos that can be observed in IceCube, the
signatures of neutrino interactions, and general methods for separating
neutrino events from the extremely large background of penetrating atmospheric
muons. Then, we present the veto techniques used to isolate a
neutrino-dominated data sample with sensitivity to neutrinos of all flavors
coming from all directions and the analysis technique that will be applied to
the sample to infer the properties of the underlying neutrino fluxes. Finally,
we present the results of the search and analysis before discussing the
implications of the result and directions for future searches.

\subsection{Sources of TeV neutrinos at Earth} 
\label{sub:sources_of_neutrinos}

Neutrinos are an inevitable byproduct of high-energy hadronic interactions that
produce weakly-decaying mesons. They may be produced either within cosmic-ray-induced
air showers, giving rise to a flux of atmospheric neutrinos
\cite{1983PhRvL..51..223G, Honda:2006,
2006JHEP...10..075G, 2007PhRvD..75l3005L, PhysRevD.86.114024,
1978PhLB...78..635B, 1983ICRC....7...22V, 1999PhLB..462..211V,
2000PhRvD..61e6011G, Martin:2003b, Sarcevic:2008, 2013AIPC.1560..350R}, or in the vicinity
of distant astrophysical accelerators, giving rise to a flux of astrophysical
neutrinos
\cite{1989cgrc.conf...21B, 1990JPhG...16.1917M, SteckerAGN, 1993PhRvD..47.5270N,
1995APh.....3..295M, 1996SSRv...75..341S, 1998PhRvD..58l3005R, 2001PhRvL..87q1102M,
2003APh....18..593M, 2005APh....23..355B, SteckerErratum, 2012ApJ...749..155E,
2000ApJ...541..707W, 2003PhRvD..68h3001R, 2006APh....25..118B, 2006PhRvL..97e1101M,
WaxmanStarburst, 2014arXiv1403.3804B, 2002ApJ...576L..33A, 2003APh....19..403G,
2008PhRvD..78f3004H, 2013arXiv1301.2437M}.

The largest contribution to the atmospheric neutrino flux comes from 2-body
decays of charged pions and kaons. These decays produce $\nu_{\mu}$ almost
exclusively because of the chiral structure of the weak interaction \cite{GaisserBook}.
IceCube has observed this flux from the northern sky
\footnote{
The ``north'' and ``south'' denote halves of the celestial sphere as observed
from the Geographic South Pole rather than terrestrial hemispheres. The majority
of atmospheric neutrinos observed
from the northern sky are produced in air showers that reach
ground level at points in the southern terrestrial hemisphere.
}
with high statistics \cite{IC59Diffuse}, and its normalization, angular distribution, and
spectral shape agree well with theoretical predictions
\cite{PhysRevD.86.114024}. Because the decays that produce $\nu_{\mu}$ also
produce muons of similar or greater energy, high-energy atmospheric $\nu_{\mu}$
from the southern sky are often accompanied by penetrating muons
\cite{Schoenert:2009}. This is a useful property that can be used to distinguish
them from astrophysical neutrinos, which are never accompanied by muons.

The flux of TeV atmospheric $\nu_{e}$ is much smaller, arising primarily from
3-body decays of $K^{\pm}$ and $K^0_L$ \cite{GaisserBook,
LipariAtmosphericLeptons}. This flux has also been observed in IceCube up to a
few TeV as an excess of events over the predicted rate from interactions of
atmospheric $\nu_{\mu}$ in the more densely-instrumented DeepCore subarray
\cite{DeepCoreCascades}. In the southern sky these too are often
accompanied by muons, though the accompanied fraction is smaller because the
muons must come from the decays of other mesons in the air
shower \cite{UncorrelatedVeto}.

A third contribution to the atmospheric neutrino flux observable in IceCube is predicted to come
from the decays of heavy, very short-lived mesons containing charm quarks like the
$D^{\pm}$. These decay preferentially to 3-body final states, with nearly equal
branching ratios to $\nu_{\mu}$ and $\nu_{e}$ \cite{PDGBook}. In addition,
their prompt decay (on the order of a picosecond) makes re-interaction
extremely unlikely, so the energy spectrum of neutrinos from charmed meson
decay follows that of the cosmic rays up to 50 PeV and is independent of the
local density of the atmosphere at production altitude \cite{GaisserBook}. This
prompt flux has not yet been conclusively observed in either muons or
neutrinos, and predictions for the normalization of the prompt atmospheric
neutrino flux vary widely \cite{Martin:2003b, Sarcevic:2008}, though more
optimistic models \cite{Naumov:1989} have already been excluded by IceCube
measurements \cite{IC59Diffuse}. Like neutrinos from light meson decays,
neutrinos from charmed meson decay are accompanied by muons. For
$\nu_{\mu}$ the accompanied fraction is similar to that in pion and kaon decays.
For $\nu_e$ it is slightly lower, because charmed meson decays are more likely
than kaon decays to produce high-energy neutrinos in relatively low-energy,
muon-poor showers \cite{UncorrelatedVeto}.

While the interactions of cosmic-ray nuclei with the Earth's atmosphere are
the largest source of high-energy neutrinos, these nuclei must come from some
population of sources. Interactions with matter or radiation fields in the
vicinity of those sources will produce neutrinos. Diffusive shock
acceleration is the most promising mechanism for accelerating protons to the
energies observed in the cosmic ray spectrum above the knee; under fairly general conditions
this is expected to produce a flux of protons with a power-law spectrum at the
source similar to $dN/dE \propto E^{-2}$ \cite{WaxmanBahcallBound, 1995PhR...258..173G,
LearnedAnnualReview, Becker2008173}. When these protons interact with other protons near the
source, they produce charged pions that decay to produce $\nu_{\mu}$ and
$\mu$ that in turn decay to produce further $\nu_{\mu}$ and $\nu_{e}$.
Long-baseline
oscillations will transform this $\nu_e:\nu_{\mu}:\nu_{\tau}$ ratio from $1:2:0$ into
approximately $1:1:1$ at Earth \cite{Learned:1995:DoubleBang, PhysRevD.80.113006}. Unless the
source environment is very dense, the energy spectrum of these astrophysical
neutrinos follows that of the progenitor protons ($\sim E^{-2}$), so their
flux exceeds that of the atmospheric neutrinos at sufficiently high energies.
Unlike atmospheric neutrinos, astrophysical neutrinos always arrive
without accompanying muons.

\subsection{Neutrino detection in IceCube} 
\label{sub:neutrino_detection_in_icecube}

The IceCube detector \cite{DAQPaper, DeepCoreDesign}
can detect neutrinos of all flavors by observing the Cherenkov photons induced
by the charged end products of charged- or neutral-current deep-inelastic
neutrino-nucleon scattering in the ice. The
signatures of these interactions fall into two broad categories: ``tracks,''
due to charged-current (CC) $\nu_{\mu}$ scattering and ``cascades,'' due to CC
interactions of $\nu_{e}$ as well as neutral-current (NC) interactions of all
flavors~\footnote{While CC $\nu_{\tau}$ interactions also produce a $\tau$
track, the brief lifetime of the $\tau$ makes it too short to be observable on
the scale of IceCube below a few tens of PeV \cite{EnergyReco}. Also, if a CC
$\nu_{\mu}$ interaction occurs inside the instrumented volume, it will appear
more track-like or more cascade-like depending on inelasticity of the
interaction and the path of the outgoing muon.}. Detections of these event
signatures and reconstruction of their properties may be used to infer the angular, energy, and flavor
distribution of an observed neutrino flux.

Roughly $1/3$ of neutrino interactions at a given energy will be
neutral-current scatterings, where the neutrino transfers on average $\sim1/3$ of its
kinetic energy to a quark in the target nucleus \cite{Gandhi1996},
producing a short ($\sim 5$~m) shower of relativistic charged particles. If the neutrino
vertex is inside or sufficiently close to the instrumented volume, the Cherenkov
radiation they induce may be detected and used to reconstruct the vertex
position and deposited energy, and to some degree the direction. The remaining
$2/3$ of interactions are charged-current scatterings that produce a charged
lepton of the same flavor as the incident neutrino in addition to a hadronic
shower. While nearly all of the neutrino's energy is converted to relativistic
charged particles, the accuracy with which the direction and energy of the
incident neutrino may be reconstructed depends strongly on the lepton flavor.
Electrons induce short electromagnetic showers whose
energy can be reconstructed to within 10\% above 10 TeV if fully inside the instrumented
volume \cite{EnergyReco}. The direction of the shower may also be inferred by
matching the spatial and temporal pattern of detected photons to a template of its angular emission
profile, though the short scattering length of the glacial ice
at the South Pole limits the angular resolution to $\sim10^{\circ}$ even above 100 TeV~\cite{EnergyReco}.
Muons lose energy much less quickly, and so can traverse kilometers of ice.
While this makes it possible to positively identify $\nu_{\mu}$ CC interactions and
reconstruct their direction to within $1^{\circ}$~\cite{IC59Diffuse,
AMANDATrackReco}, it is more difficult to infer the initial neutrino energy when
only a segment of the muon track can be observed.

\begin{figure}[htb]
	\centering
		\includegraphics[scale=1]{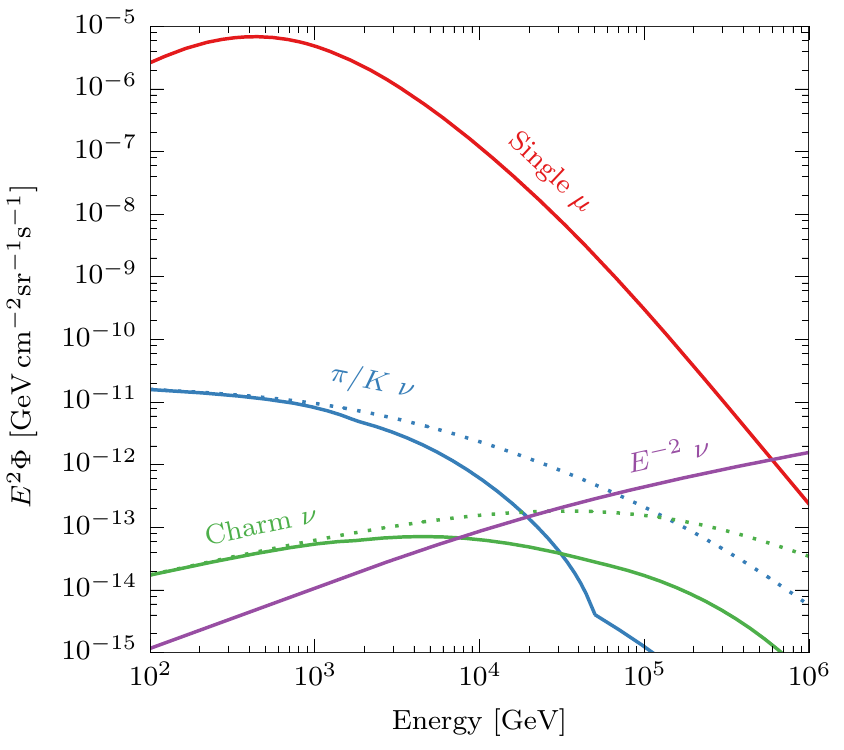}
	\caption{
	Fluxes of vertically-downgoing muons and neutrinos detectable in IceCube.
The upper line shows the flux of penetrating, single atmospheric muons at the
depth of IceCube, while the remaining lines show neutrino fluxes multiplied by
the probability that a neutrino of the given energy would interact in 1 km of
glacial ice. The dotted lines show the total interacting flux of atmospheric
neutrinos of all flavors \cite{Honda:2006, Sarcevic:2008}, while the corresponding solid lines
show the interacting flux that arrives at the depth of IceCube without
accompanying muons above 1 TeV \cite{UncorrelatedVeto}. Accompanying muons suppress the effective
$\nu_{\mu}$ flux from $\pi$ and $K$ decay below the level of the effective $\nu_{e}$ flux
from $K$ decay at 50 TeV, producing a kink in the spectrum. The $E^{-2}$ astrophysical neutrino flux, shown here
with the normalization of \cite{HESE_3year}, always arrives without
accompanying muons.
	}
	\label{fig:underground_fluxes}
\end{figure}

The background to neutrino searches comes from atmospheric muons that penetrate
the 1.5 km of ice overburden; they outnumber neutrino interactions by a factor
of $\sim 10^6$. The traditional method of separating
neutrino events from the penetrating muon background is to select only events
where a muon comes from below the horizon, where the bulk of the Earth
completely absorbs atmospheric muons. While this makes the effective
volume of the detector much larger than its geometrical volume, the method is
only sensitive to CC $\nu_{\mu}$ interactions and cannot be extended much more
than $\sim5^{\circ}$ above the geometric horizon before penetrating muons begin to
overwhelm neutrino-induced ones. An alternative approach is to select events
that start inside the instrumented volume: neutrinos are completely invisible,
whereas incoming muons induce Cherenkov photons that should be detected as the
muon enters the fiducial volume. Such a veto-based event selection can be
sensitive to neutrinos of all flavors from all directions, though it must
sacrifice some effective volume to implement the veto. Since a muon veto also
rejects down-going atmospheric neutrinos that are accompanied by muons, a veto-based event selection
also enjoys a smaller atmospheric neutrino background than the traditional
up-going neutrino selection \cite{Schoenert:2009}, as illustrated in
Fig.~\ref{fig:underground_fluxes}.

\section{Event selection} 
\label{sec:event_selection}

\begin{figure*}[htb]
	\centering
		\includegraphics[scale=1]{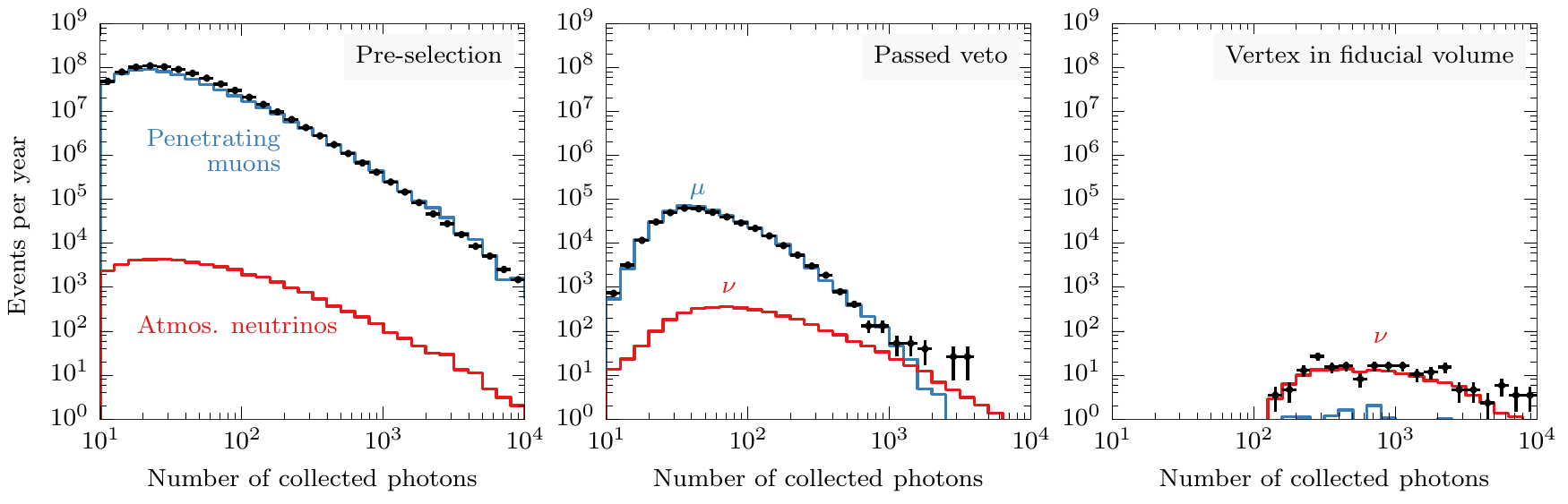}
	\caption{
	Distribution of photon counts per event after each stage of the event selection.
The total number of collected photons is on average proportional to the total
deposited energy; for example, $10^3$ photons correspond to roughly 10 TeV
deposited energy. The stepped lines show the
prediction from Monte Carlo simulation of penetrating atmospheric muons (blue)
atmospheric neutrinos (red), while the points show experimental data. Left:
Pre-selected events transmitted from the South Pole
(Sec.~\ref{sub:pre_selection}). Center: Removed events with veto hits
(Sec.~\ref{sub:veto}). Right: Fiducial volume scaled with photon count
(Sec.~\ref{sub:fiducial_volume_scaling}).
	}
	\label{fig:cut_progression}
\end{figure*}

Active vetoes have been used in IceCube to isolate neutrino
interactions inside the instrumented volume of the detector before. The method
has been used effectively in the DeepCore subarray \cite{DeepCoreCascades} and in
the full detector above 100 TeV \cite{HESE, HESE_3year}\footnote{Veto methods
were also used in previous searches for cascade events in AMANDA and IceCube
\cite{AMANDACascades, AMANDACascades5Year, IC22Cascades, IC40Cascades}, but
were not the primary method of background rejection.}. However, it has not been
explored in the region between 10 and 100 TeV, where neutrinos from charmed meson decay in the atmosphere
should be observable. Extending veto methods to this intermediate energy range without
sacrificing sensitivity to all neutrino flavors or to neutrinos from the
southern sky requires new techniques. Figure~\ref{fig:cut_progression} illustrates the effects of the
three stages of the event selection that will be described in this section: a pre-selection cut to reduce the data rate to a manageable
level, veto cuts to remove events with signs of incoming muons, and a
fiducial volume cut to remove events where incoming muons cannot be vetoed with
sufficiently high probability.

\subsection{Data collection and pre-selection} 
\label{sub:pre_selection}

The IceCube detector \cite{DAQPaper, DeepCoreDesign} consists of 5160 Digital
Optical Modules (DOMs) buried in the glacial ice at the South Pole, instrumenting
a total volume of approximately 1~km$^3$. The DOMs are attached
to cables that provide power and communication with the data acquisition system
on the surface of the glacier. Each of these ``strings'' hosts 60 DOMs; 78 of
the strings are spaced 125 m apart on a hexagonal grid with DOMs placed every
17 m from 1450 to 2450 m below the surface, while the remaining 8 strings form
the DeepCore in-fill array \cite{DeepCoreDesign}. These in-fill strings are
30--60 m from the nearest string with 50 DOMs placed every 7 m between 2100 and
2450 m below the surface, where the glacial ice is most transparent, and 10
DOMs placed every 10 m between 1750 and 1850 m below the surface. The data that
will be presented in Sec.~\ref{sec:results} were taken with the nearly-complete
79-string detector configuration from May 2010 to May 2011 and the first year
of the complete 86-string detector from May 2011 to May 2012.

Each DOM consists of a 25 cm diameter photomultiplier tube (PMT) \cite{PMTPaper}, power supply,
and digitization electronics housed in a borosilicate glass pressure sphere.
The PMT signal is digitized and stored for transmission to the surface whenever
the PMT output current exceeds 1/4 of the mean peak current of the pulse
amplified from a single photo-electron (PE); if a neighboring or next-to-neighboring DOM on the same
string also triggers within 1 $\mu$s (local coincidence) the readout extends
for 6.4 $\mu$s, otherwise the readout only includes a 75 ns window around the
peak current in the first 400 ns after the local trigger. The digitized
waveforms are transmitted to the surface, where they are assembled into events
by a software trigger \cite{DAQPaper, String21Paper}.

The arrival times of individual photons and photon bunches are reconstructed by
deconvolving the characteristic single-PE pulse shape from the digitized
waveforms \cite{EnergyReco}, and the resulting times and photon counts are used to
reconstruct the vertices, directions, and energies \cite{AMANDATrackReco,
EnergyReco} of the relativistic charged particles that induced the detected
Cherenkov photons.

The first stage of the event selection is done at the South Pole to reduce the
data rate enough for transmission over a satellite link; the resulting sample, shown
in the left panel of Fig.~\ref{fig:cut_progression}, includes $\sim1$\% of
all triggering events. While this selection retains the majority of
triggering neutrino events, they are still out-numbered 10,000 to 1 by
penetrating muons. The remaining selection steps remove this background,
leaving a nearly-pure sample of neutrino events.

\subsection{Veto} 
\label{sub:veto}

This selection targets interactions of isolated neutrinos that occur inside the instrumented
volume, producing a cascade at the neutrino vertex as well as an out-going track
in the case of CC $\nu_{\mu}$. These can be distinguished from incoming muons
(and atmospheric neutrinos accompanied by muons) by photons detected before the putative neutrino
vertex. The simplest approach is the one employed in \cite{HESE}, where the
outermost layer of PMTs is used as an active veto. The event is rejected if photons are detected on
the outermost layer before the estimated vertex time. Spurious vetoes due to detector noise are mitigated by requiring that
the photons are detected in local coincidence \cite{DAQPaper} at times that are
causally compatible with the estimated vertex. For very bright events with thousands of
detected photons, this single cut is sufficient to suppress the muon background
below the level of atmospheric neutrinos (Fig.~\ref{fig:cut_progression},
center panel).

\begin{figure*}[htb]
	\centering
	\subfloat[][
		Penetrating muon before its largest energy loss.
		The dashed grey lines mark the positions at
		which photons induced by a muon would be detected with minimal and maximal
		delay. The photon that falls inside this window is counted towards the veto
		total, while the random noise photon that falls outside the window is not.
	]{
		\includegraphics[width=0.32\textwidth]{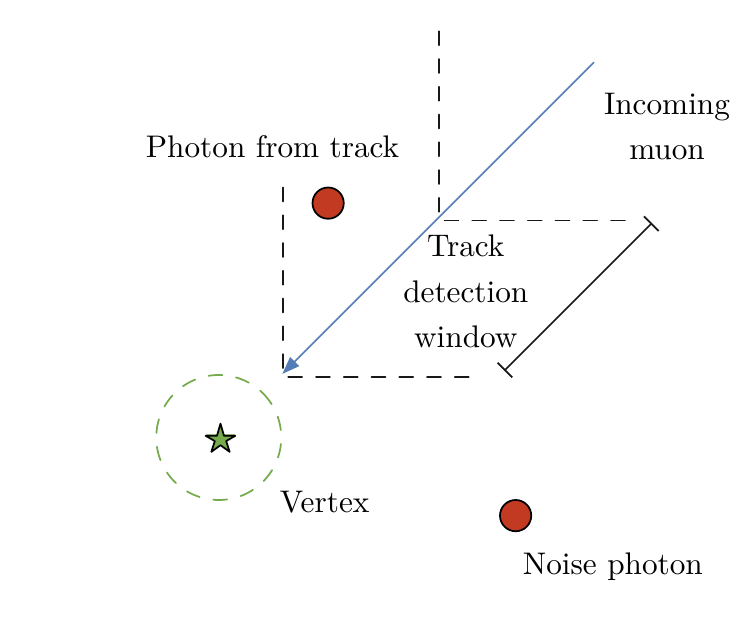}
		\label{fig:veto_sketch:incoming_before}
	}
	\hspace*{\fill}
	\subfloat[][
		Penetrating muon after its largest energy loss.
		The dashed circle marks the positions where photons propagating
		from the vertex at the speed of light in ice would be detected with minimal
		delay. Here the photon is not counted towards the veto since it is detected
		at a time compatible with propagation from the reconstructed vertex.
	]{
		\includegraphics[width=0.32\textwidth]{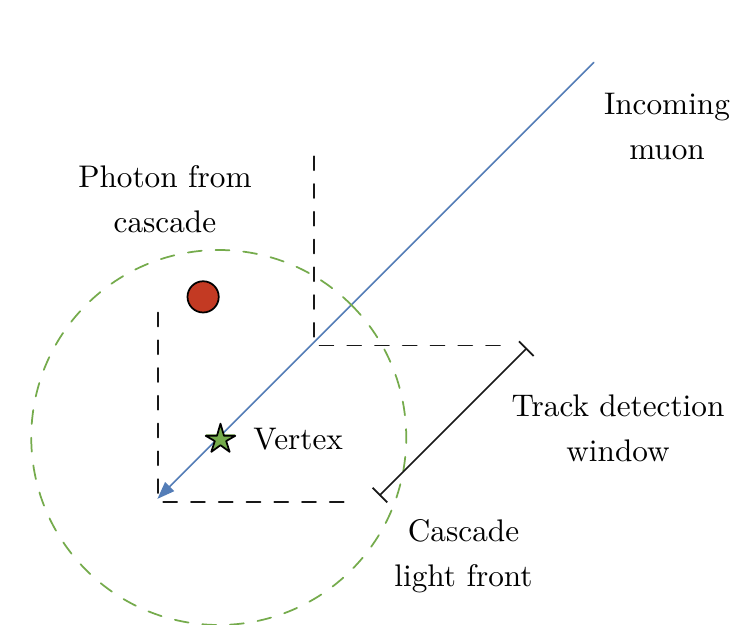}
		\label{fig:veto_sketch:incoming_after}
	}
	\hspace*{\fill}
	\subfloat[][
		Neutrino-induced muon.
		Photons induced at the cascade vertex spread outwards at the speed of light
		in ice, while the muon moves at the speed of light in vacuum. Eventually the
		muon out-runs the light front from the cascade, and photons collected in the
		track detection window can be used to positively
		identify an out-going muon in the event.
	]{
		\includegraphics[width=0.32\textwidth]{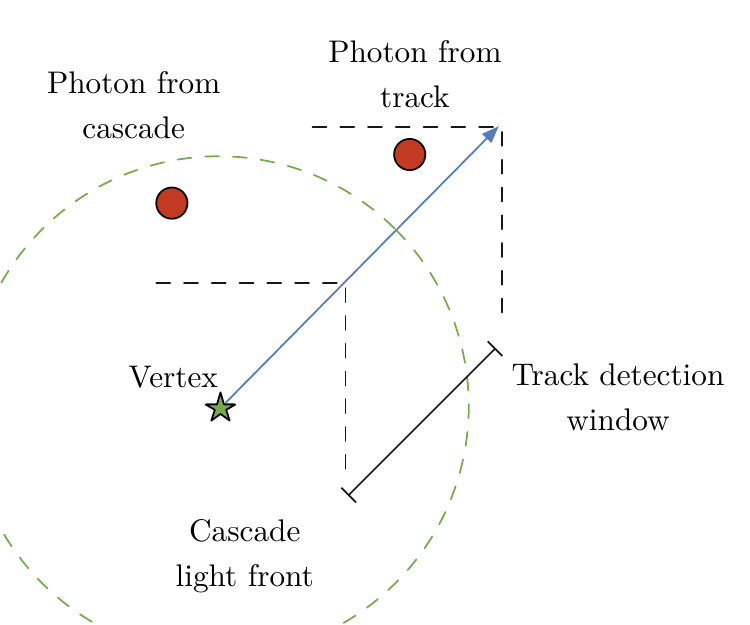}
		\label{fig:veto_sketch:starting}
	}
	
	\caption{An illustration of the incoming-muon veto procedure described in Sec.~\ref{sub:veto}. Each panel
	shows a snapshot in time with the current position of the muon marked by the
	blue arrowhead and the position of the reconstructed vertex marked by a green
	star. \protect\subref{fig:veto_sketch:incoming_before} shows a penetrating muon
	before its largest energy loss with a photon detection that counts towards the veto,
	while \protect\subref{fig:veto_sketch:incoming_after} shows the same configuration
	after the largest energy loss with an ambiguous photon detection that does
	not count towards the veto. 
	\protect\subref{fig:veto_sketch:starting} shows how the technique can be
	inverted to detect starting tracks.}
	
	\label{fig:veto_sketch}
\end{figure*}

As the energy threshold is lowered, the number of background muons increases
rapidly, while their average energy loss rate decreases, a combination that
overwhelms the ability of the single layer to reject incoming muons. In order
to extend the selection to lower energies, a second kind of veto, similar to those
employed in \cite{IC79SolarWIMPPRL, DeepCoreCascades}, is required.
The first modification is to remove the requirements that veto photons be detected
on the outermost layer of PMTs and in local coincidence. This allows isolated photon
detections anywhere in the instrumented volume to veto an incoming track, which lowers
the energy threshold but also loses signal events to spurious vetoes caused by
noise. In order to mitigate the signal loss, a second modification is required:
photons are only considered for
veto if they are detected at a time and position consistent with an incoming
track but inconsistent with the reconstructed cascade vertex, as shown in
Fig.~\ref{fig:veto_sketch:incoming_before}.

The most resilient background comes from single muons with a single
disproportionately large stochastic energy loss. Such events can be completely dominated
by the photons induced by the largest energy loss, causing track reconstruction
algorithms that assume uniform light emission to fail to find the correct direction.
The position and time of the vertex can, however, be reconstructed reliably
regardless of the presence of a muon track. In order to ensure that veto
photons will be found, the search is repeated for each of 104 different down-going
track hypotheses\footnote{The directions are chosen from the upper hemisphere of a HEALpix \cite{HEALpix} grid.}
that pass through the reconstructed vertex.
The track hypothesis with the
largest number of associated veto photons is considered the best. A photon
is associated with an incoming track if it is detected
\begin{itemize}
	\item at least 50 ns before the earliest possible time for a photon induced
	      at the previously reconstructed vertex,
	\item between 15 ns before and 1000 ns after the earliest possible time for
	      a photon induced by the hypothetical muon,
	\item and no more than 100 m from the hypothetical muon trajectory.
\end{itemize}
The event is rejected if the best track has more than two associated photons.

\subsection{Fiducial volume scaling} 
\label{sub:fiducial_volume_scaling}

\begin{figure}[htb]
	\centering
		\includegraphics[scale=1]{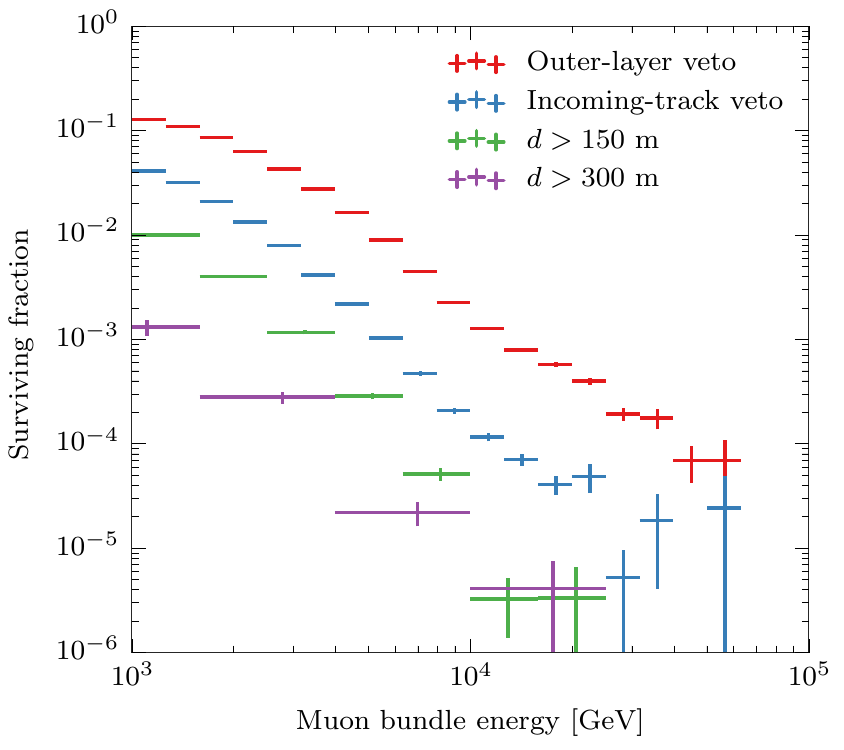}
	\caption{
	Fraction of pre-selected penetrating muon background events
(Sec.~\ref{sub:pre_selection}) that pass the veto conditions
(Sec.~\ref{sub:veto}), derived from MC simulation. The outer-layer veto reduces
the rate of the highest-energy muons by $10^4$, but degrades rapidly at
lower energies. The incoming-track veto scales in a similar way with respect to
energy, but is more sensitive because it considers isolated photon detections.
In contrast to the outer-layer veto, its efficiency also improves with
increasing distance $d$ from the detector border of the reconstructed vertex.
	}
	\label{fig:veto_passing_rate_l2}
\end{figure}

\begin{figure}[htb]
	\centering
		\includegraphics[scale=1]{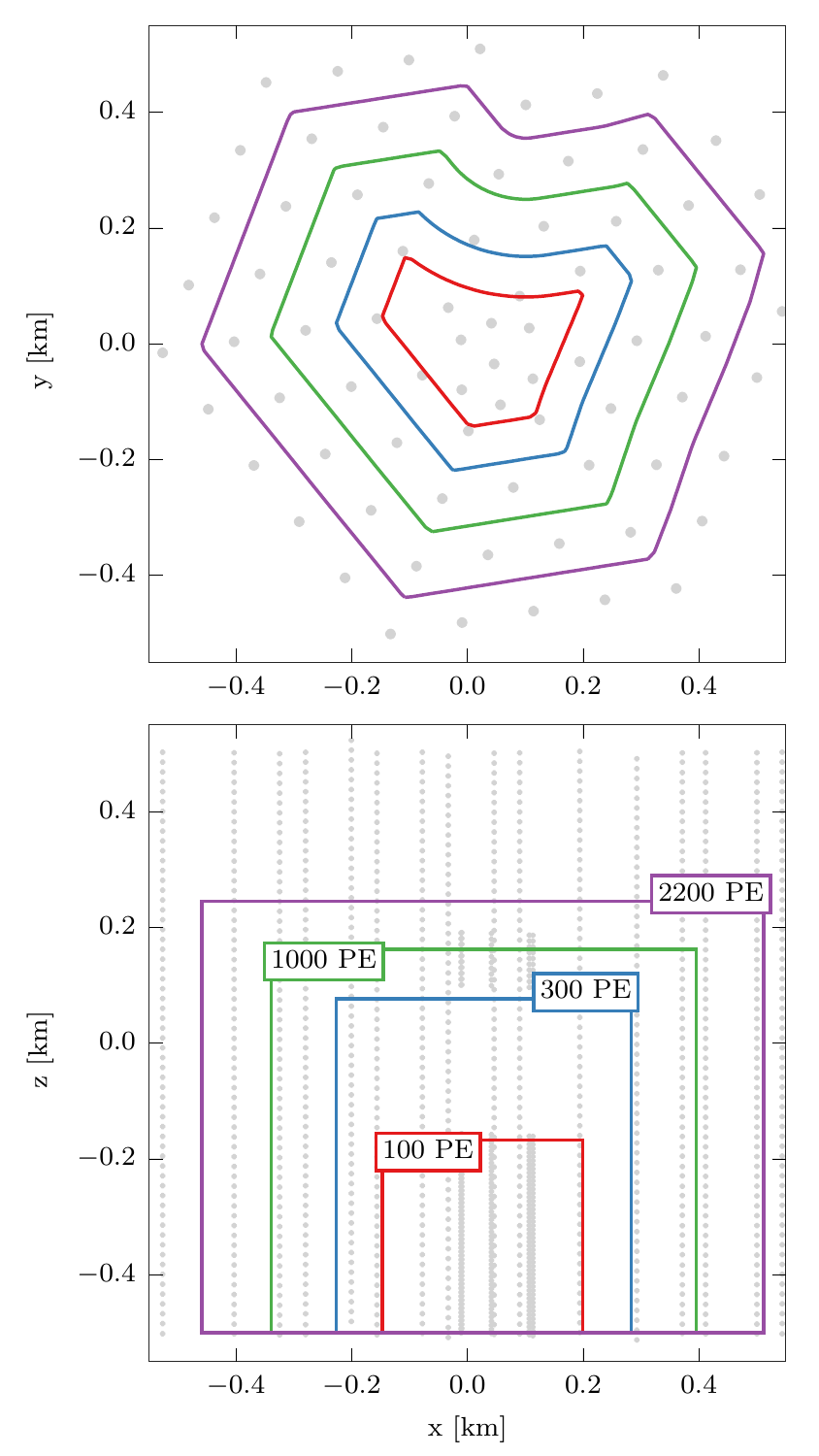}
	\caption{Fiducial volume scaling function evaluated at four different photon counts.
	Top: Overhead view, showing the positions of the IceCube strings and the
	boundaries of the fiducial volume for events with a given total photon count.
	Bottom: Side view, showing the modules along strings.}
	\label{fig:volume_scaling}
\end{figure}

Since the effectiveness of the track-based veto is proportional to the
probability of detecting at least 2 photons from an incoming muon before the
reconstructed vertex, it increases in proportion to the number of detectable
photons the muon induces and the number of PMTs it passes on its way to the
vertex as shown in Fig.~\ref{fig:veto_passing_rate_l2}. This relationship can be exploited to maintain sufficient penetrating
muon rejection at low energies by requiring a minimum distance between the
reconstructed vertex and the edges of the instrumented volume that increases as
the number of collected photons decreases as shown in
Fig.~\ref{fig:volume_scaling}. At the lowest photon counts the fiducial
volume is reduced to the DeepCore subarray with the remainder of the detector
used as a veto; as the photon count increases the veto reduces to the outermost
layer of PMTs as in \cite{HESE}. This selection creates the neutrino-dominated
sample shown in the right panel of Figure~\ref{fig:cut_progression}. Tables of
the effective area of this selection as a function of incident neutrino energy
and zenith angle are provided in the online supplemental materials for this
article.

\section{Analysis method} 
\label{sec:analysis_method}

The events that pass the selection defined above arise from four sources:
conventional atmospheric neutrinos, prompt atmospheric neutrinos, penetrating
atmospheric muons, and high-energy extraterrestrial neutrinos. Each of these
components produces a distinct distribution in the observables chosen for this
analysis: reconstructed deposited energy, reconstructed zenith angle, and the
presence or absence of a detectable outgoing muon track. We disentangle the
contributions of each of these components by fitting a model of their
observable distributions predicted from Monte Carlo simulation to the observed
data. In this section we define the observables and discuss their expected
distributions for each component. Then, we present the likelihood fitting
technique and the method for determining statistical errors on the fit
parameters.

The deposited energy is reconstructed by fitting the observed spatial and temporal
distribution of detected photons to a template derived from simulations of
single, point-like electromagnetic cascades as described in \cite{EnergyReco}.
This electromagnetic-equivalent energy can be resolved to within 10\% (68\% C.L.) and is a
proxy for the neutrino energy; for CC $\nu_e$ it is a nearly unbiased estimator of the
neutrino energy \cite{Beacom:2004:ShowerPower}, while for all other interaction types it is on average
proportional to the neutrino energy. Since the template depends on the
orientation of the cascade with respect to the DOM, the same technique yields a
direction as well, with a typical zenith angle resolution of 15\degree (68\% C.L.) in the sample presented here.
Outgoing muon tracks are identified by inverting the incoming-track veto of
Sec.~\ref{sub:veto} as shown in Fig.~\ref{fig:veto_sketch:starting} and
classifying events with more than 10 photons in the detection window of an
outgoing track as track-like. 35\% (60\%) of CC $\nu_\mu$ events from the
conventional atmospheric ($E^{-2}$ astrophysical) spectrum satisfy this criterion;
in the remaining events, the outgoing muon escapes the instrumented volume without
being detected, and the event is mis-classified as a cascade. The reverse case
is much rarer: 0.001\% (3\%) of NC $\nu_\mu$ events from the
conventional atmospheric ($E^{-2}$ astrophysical) spectrum are mis-classified as
tracks. For track-like events the zenith angle is taken
from the best-fit outgoing track, as the large displacement of the associated photons
from the neutrino interaction vertex provides a better constraint than the
initial cascade. Explicitly retaining these track-like events, rather than
rejecting them with cuts designed to select cascade events, provides a built-in
control sample that we use to check our understanding of the selection's neutrino acceptance.

\begin{figure*}[htb]
	\centering
		\includegraphics[scale=1]{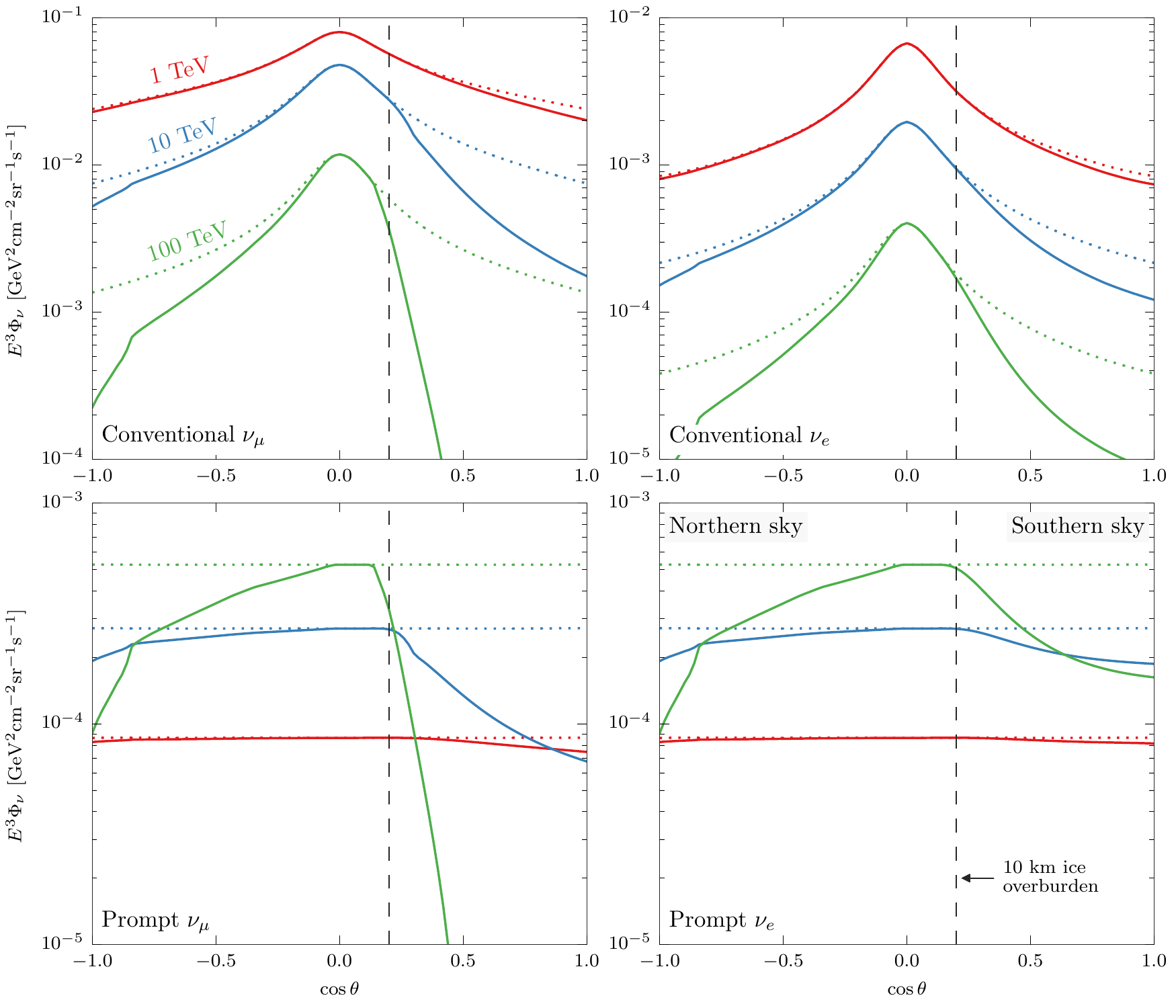}
	\caption{
	Atmospheric neutrino flux models used in this analysis. The dotted lines in
each panel show the neutrino fluxes at Earth's surface as a function of true zenith
angle at 1, 10, and 100~TeV. The conventional fluxes are taken from
\cite{Honda:2006} and the prompt flux from \cite{Sarcevic:2008}; both were
corrected to account for the cosmic ray flux of \cite{Gaisser2012801}. The
solid lines show the fluxes of $\nu_\mu$ and $\nu_e$ that can be observed as
isolated neutrino interactions in IceCube. The observable fluxes are suppressed
in the northern sky ($\cos\theta \leq 0.2$, to the left of the vertical dashed line) by absorption in the Earth,
especially in its much denser core ($\cos\theta<-0.8$) \cite{Gandhi1996},
 and in the southern sky ($\cos\theta>0.2$, to the right of the line) by
self-veto by accompanying muons
\cite{UncorrelatedVeto}.
Astrophysical neutrinos are absorbed in the Earth as
well, but are never accompanied by muons.
	}
	\label{fig:atmospheric_fluxes}
\end{figure*}

Conventional atmospheric neutrinos are produced in the decays of charged pions
and charged and neutral kaons in the atmosphere. Since these mesons are
relatively long-lived, they are more likely to re-interact and lose energy than
to decay to produce neutrinos. This competition affects both the deposited
energy and the zenith angle distributions. The neutrino energy spectrum above 1 TeV
is one power in energy steeper than that of the input cosmic ray spectrum,
causing the deposited energy distribution to peak at low energies. The flux
is largest at the horizon where the average density of the atmosphere along the
air shower axis is smallest \cite{LipariAtmosphericLeptons}, causing the
reconstructed zenith angle distribution to peak at the horizon as well. In the
southern sky these effects are further enhanced by accompanying muons
produced in the same air shower: they trigger the veto, removing high-energy,
down-going atmospheric neutrino events from the sample \cite{Schoenert:2009}.
The veto removes $\nu_{\mu}$ more efficiently than $\nu_e$, reducing the
observable conventional atmospheric neutrino flux to the sub-dominant $\nu_e$
component at sufficiently high energies and small zenith angles.
In this analysis, the flux model for conventional atmospheric neutrinos is taken
from a parametrization of the calculation of \cite{Honda:2006}, corrected to
account for the cosmic ray flux of \cite{Gaisser2012801} as described in \cite{IC59Diffuse} and for the fraction
of atmospheric neutrinos that are vetoed \cite{UncorrelatedVeto} by accompanying
muons. These fluxes are shown in the upper two panels of
Fig.~\ref{fig:atmospheric_fluxes}.

Atmospheric neutrinos produced in the decays of charmed mesons behave
differently. The lifetimes of charmed mesons are extremely short, so they nearly always decay
promptly before re-interacting, producing an isotropic neutrino flux with nearly the
same spectral index as that of the primary cosmic rays. Since neutrinos with
accompanying muons are vetoed in the event selection, the observable flux is
depleted in the southern sky. The overall suppression is weaker than for conventional
atmospheric neutrinos because of the larger fraction of $\nu_e$. The flux model for prompt atmospheric
neutrinos is taken from a parameterization of the calculation of
\cite{Sarcevic:2008} with corrections for the cosmic ray flux and veto passing
fraction. These fluxes are shown in the lower two panels of
Fig.~\ref{fig:atmospheric_fluxes}.

The events that pass the final selection include a small but nearly irreducible
background of penetrating atmospheric muons that go undetected before
depositing a large fraction of their energy in the glacial ice in a single, catastrophic loss.
These events come exclusively from the southern sky because muons
cannot penetrate the bulk of the Earth, and are sharply peaked at the
deposited-energy threshold of the selection because the veto removes muons with
increasing efficiency at higher energies. The flux model for penetrating
atmospheric muons is taken from a parametrization of
\textsc{corsika}~\cite{CORSIKA} air-shower simulations with the cosmic ray flux
parameterization of \cite{Gaisser2012801}, using the sum of muons from the
decays of light hadrons predicted in \textsc{sibyll}~\cite{SIBYLL_2.1} and from
the decays of charmed hadrons predicted in \textsc{dpmjet}~\cite{DPMJETCharm}
to obtain an upper bound on the underground flux of single muons. This combined
model predicts a total of 14 penetrating muon events in the sample.

Overlaid on these atmospheric components is a flux of high-energy neutrinos
of astrophysical origin.
Their energy distribution is harder than those of any of the other sources of
neutrinos, and we found that they are the dominant source of events with more than 100 TeV
deposited energy \cite{HESE, HESE_3year}. The Earth absorbs a significant fraction of upward-going
neutrinos above 100 TeV \cite{Gandhi1996}, so the highest-energy of these are
concentrated around the horizon and in the southern sky. Since the
sources of these neutrinos are unknown, the shape of their energy and
angular distribution cannot be predicted exactly, and given the limited number
of neutrino events that can be detected, only very simple models can be tested.
Neutrinos associated with the extragalactic sources of the highest-energy cosmic
rays are assumed to be isotropically distributed and follow a power law
energy distribution of approximately $E^{-2}$ \cite{WaxmanBahcallBound} and arrive
at the Earth as equal parts $\nu_e$, $\nu_{\mu}$, and $\nu_{\tau}$ due to oscillations
\cite{PhysRevD.80.113006}. We parameterize the diffuse astrophysical neutrino flux as
\begin{equation}
	\label{eq:powerlaw}
	\Phi_{\rm astro} = \Phi_{0}
	\left(\frac{E}{E_0} \right)^{-\gamma} ,
\end{equation}
where $\Phi_{0}$ is the $\nu + \overline{\nu}$ flux for each
flavor at $E_0=10^5$~GeV in units of $\rm{GeV^{-1} \, cm^{-2} \, sr^{-1} \, s^{-1}}$ and $\gamma$ is the spectral index\footnote{
Equivalently, $E^2 \Phi_{\rm astro} =  \Phi_0 \times (10^{10} \,\,{\rm GeV}^2) \times \left(\frac{E}{E_0}\right)^{2-\gamma} $}.
More generally, $\gamma$ can be
allowed to vary to account for the spectra expected from specific classes of
sources, for example TeV photon emission from active galactic nuclei ($2.2
\lesssim \gamma \lesssim 2.6$) \cite{Becker2008173} or interactions of cosmic
rays with dense gas clouds while magnetically confined in starburst galaxies
$2.0 \lesssim \gamma \lesssim 2.25$) \cite{WaxmanStarburst}.

\begin{figure*}[htb]
	\centering
		\includegraphics[scale=1]{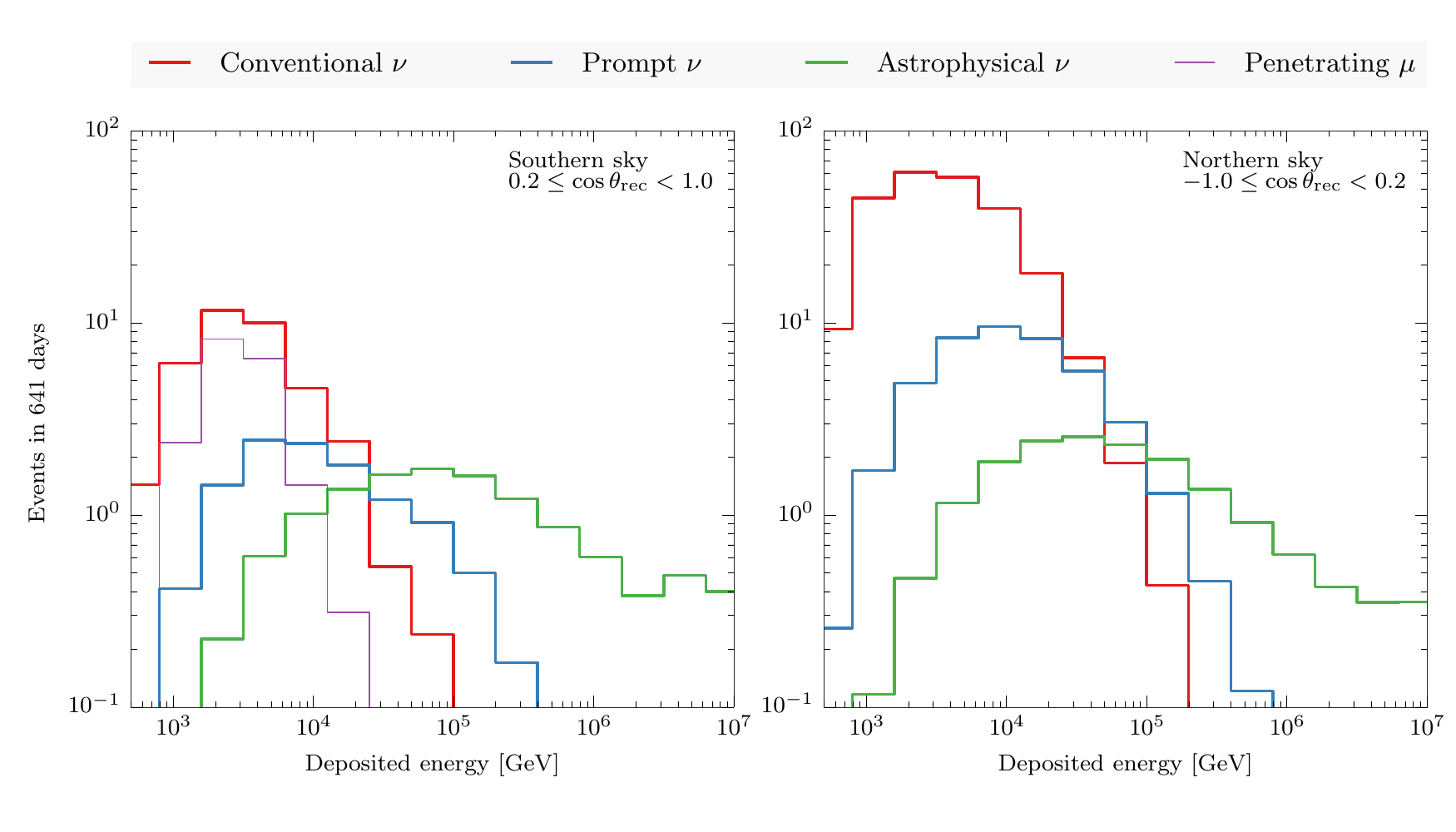}
	\caption{
	Average deposited-energy spectra expected from the various sources of
neutrinos in this analysis from the southern and northern skies. The
conventional atmospheric component corresponds to the calculation of
\cite{Honda:2006}, with corrections for the knee of the cosmic-ray spectrum and
the fraction vetoed by accompanying muons, while the prompt component
corresponds to the calculation of \cite{Sarcevic:2008} with similar
corrections, but with the normalization taken from the previously-published upper limit of
3.8 \cite{IC59Diffuse}. The astrophysical component corresponds to
Eq.~\eqref{eq:powerlaw} with $\Phi_0=10^{-18} \, {\rm {GeV^{-1} \, cm^{-2} \, sr^{-1} \, s^{-1}} }$ and $\gamma=2$.
	}
	\label{fig:energy_north_south_demo}
\end{figure*}

Figure~\ref{fig:energy_north_south_demo} shows the average deposited energy
distributions expected from the northern and southern skies from the sources of
neutrinos presented above. For the purposes of this analysis the southern sky
extends only to $80\degree$ instead of the geometric horizon, since the $\sim10$
kilometers water-equivalent of overburden at this zenith angle are already
sufficient to remove the vast majority of atmospheric muons that would
otherwise veto atmospheric neutrino events (see Fig.~\ref{fig:atmospheric_fluxes}). Conventional atmospheric neutrinos
are concentrated at deposited energies of a few TeV and in the northern sky
around the geometric horizon; the smaller contribution above the horizon in the
southern sky is further suppressed by vetoing muons (see
Fig.~\ref{fig:atmospheric_fluxes}). Since conventional atmospheric neutrinos have the largest fraction of
$\nu_{\mu}$, they represent the largest contribution to the track-like portion
of the data sample. The fraction of events in the track sample provides another
constraint on the conventional atmospheric
flux in addition to the deposited-energy and zenith distributions. The astrophysical
component, shown as an isotropic flux with a normalization at the best fit of \cite{HESE_3year},
dominates above 100 TeV in the northern and
30 TeV in the southern sky. The prompt atmospheric component, shown with a
normalization at the previously-published upper limit of 3.8 \cite{IC59Diffuse}, never provides a
dominant contribution to the observed event rate. Instead, a large prompt
component appears as an excess over the conventional atmospheric and
astrophysical components in the northern sky in the 30--60 TeV region that is
not matched in the southern sky. While the exact cross-over energies depend on
the normalization and spectral index of the astrophysical component, the ordering
of the energy ranges where each component can be constrained is generic.

The parameters of the model that best fit the observed data are determined
through a binned likelihood fit. In this procedure, the data
sample is binned in the three observables: reconstructed deposited energy,
reconstructed zenith angle, and presence of a detectable outgoing track. The
observed count $n_i$ in each bin $i$ is compared to a model that
predicts the mean count rate $\lambda_i$ in each bin through a Poisson likelihood function

\begin{equation}
	\label{eq:poisson_likelihood}
	L = \prod_{\mathrm{bins} \, i} \frac{e^{-\lambda_i} \lambda_i ^ {n_i}}{ n_i !} .
\end{equation}

The mean rates $\lambda_i$ are computed by applying the event selection to
simulated data and binning the observables of the surviving events in the same
way as the experimental data, weighted according to the set of flux models
under consideration. This convolves the flux model with the detector response
to obtain observable distributions, a procedure known as forward folding. The
model is fit to the data by varying its parameters until
\eqref{eq:poisson_likelihood} is maximized.

The 68\% confidence ranges on each model parameter are obtained from a likelihood-ratio test.
To test whether a value of one parameter can be
rejected at the desired confidence level, the parameter is constrained to that
value while all other parameters are varied to maximize the conditional likelihood.
The ratio between this conditional likelihood maximum and the global maximum is the profile likelihood.
It can be used to construct a test statistic
\begin{equation}
	\label{eq:test_statistic}
	-2\Delta \ln L = -2(\ln L - \ln L_{\rm max})
\end{equation}
whose distribution approaches that of a $\chi^2$ with 1 degree of freedom in
the large-sample limit \cite{wilks1938}.
If $-2\Delta \ln L > 1 \,\, (2.71)$, then the
tested value of the model parameter is rejected at more than 68\% (90\%) confidence
\footnote{
The test statistic does not necessarily follow a $\chi^2$ distribution when the
sample size is finite or a parameter is close to a bound. In such cases
the exact confidence level can be derived numerically from Monte Carlo trials.
In this analysis, however, the exact confidence intervals were found to be only
slightly smaller than intervals derived from the $\chi^2$ approximation.
}
.

\section{Results} 
\label{sec:results}

283 cascade and 105 track events passed the final selection criteria in 641
days of data-taking. Of those 388 events, 106 deposited more than 10 TeV at the
cascade vertex, and 9 deposited more than 100 TeV. At high energies the
selection overlaps nearly completely with the selection of \cite{HESE}: 7 of
the 9 events depositing more than 100 TeV were also in the previous selection.

\begin{figure*}[htb]
	\centering
		\includegraphics[scale=1]{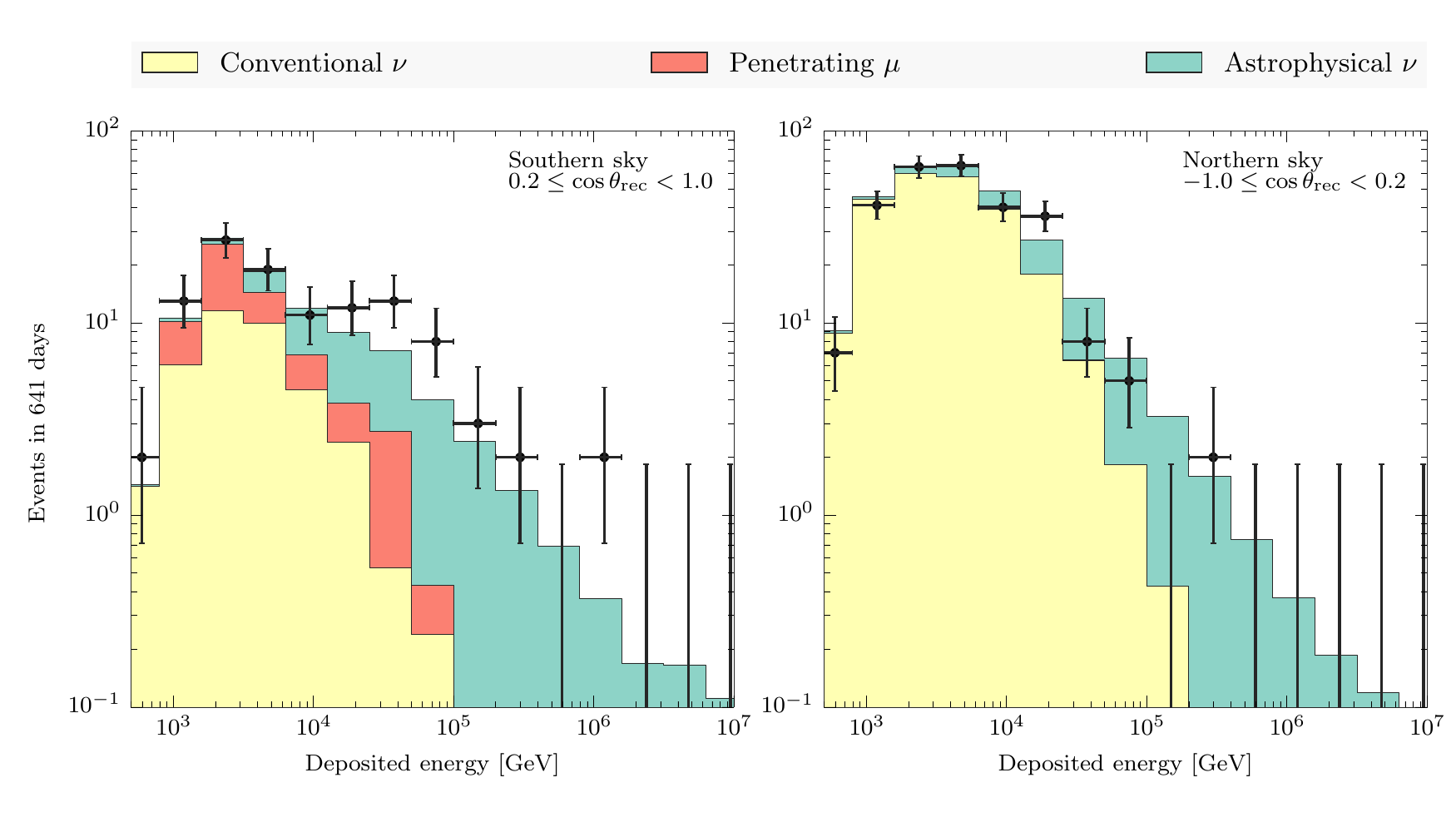}
	\caption{
	Deposited energy spectra from the northern and southern skies (points) with
the best-fit combination of atmospheric and astrophysical contributions from
Table~\ref{tab:bestfit}. Below 3 TeV, the events observed from the northern sky
are adequately explained by conventional atmospheric neutrinos. In the same
energy range in the southern sky, penetrating atmospheric muons account for the
remaining events. Above 10 TeV, an extra component is required to account for the
observed high-energy events, especially those in the southern sky. Since atmospheric
neutrinos of any kind are often vetoed by accompanying muons, the excess is best
explained by astrophysical neutrinos. We interpret the excess over the best-fit
sum around 30 TeV as a statistical fluctuation.
	}
	\label{fig:energy_north_south}
\end{figure*}

\begin{figure}[htb]
	\centering
		\includegraphics[scale=1]{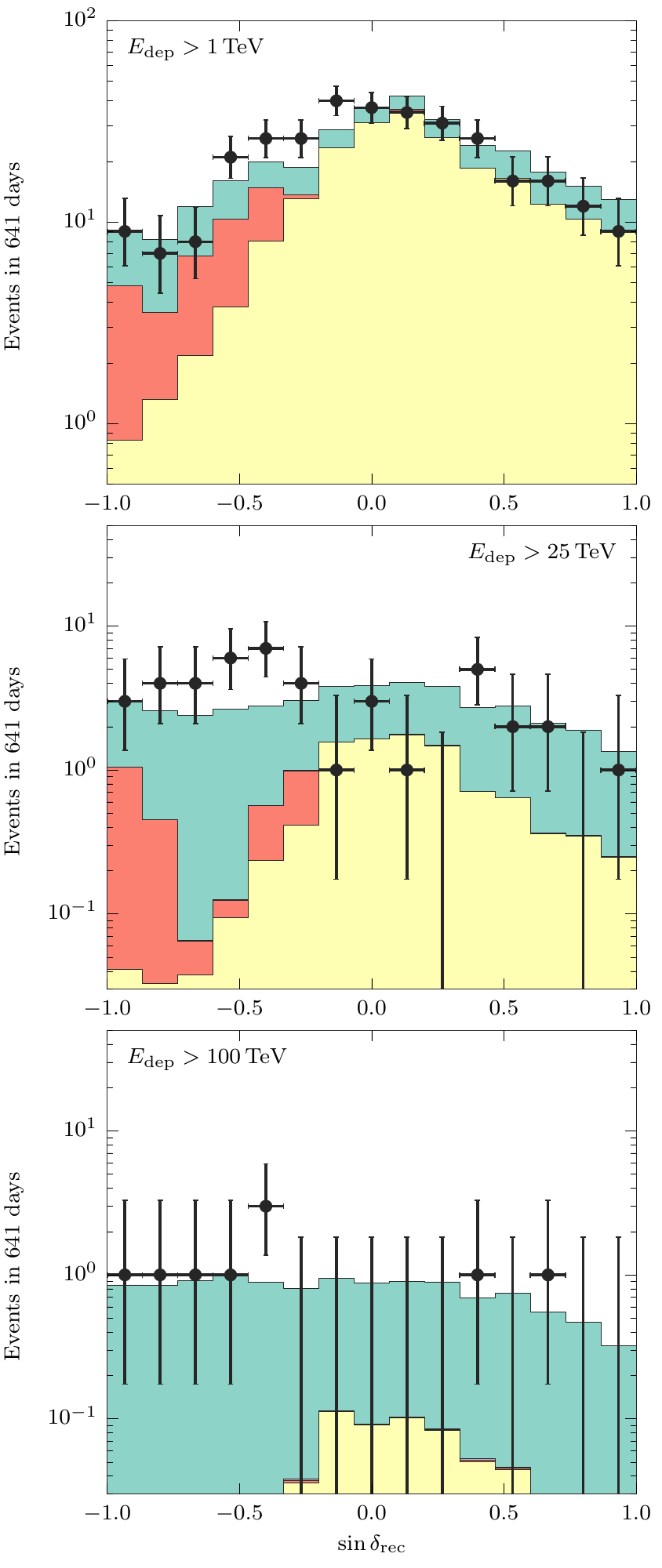}
	\caption{Zenith angle distribution of events depositing more than 1, 25,
	and 100 TeV (points) with the best-fit combination of atmospheric and astrophysical
	contributions from Table~\ref{tab:bestfit}, using the same color scheme as
	in Fig.~\ref{fig:energy_north_south}. At the lowest energies the sample
	is concentrated at the horizon, as expected from conventional atmospheric neutrinos.
	The astrophysical component contributes significantly to the sample above
	25 TeV, and the bulk of the sample is down-going. By 100 TeV only the astrophysical
	component remains, and the up-going flux is suppressed by absorption in the Earth.
	}
	\label{fig:zenith}
\end{figure}

The likelihood-fit approach described above was used to
determine the fluxes of neutrinos and muons compatible with the observed
events. In the first fit, the normalizations of the penetrating atmospheric
muon component, the conventional and prompt atmospheric neutrino components, as well as the
per-flavor normalization $\Phi_0$ and spectral index $\gamma$ of the astrophysical component
were allowed to vary freely, resulting in the best-fit parameters shown in
Tab.~\ref{tab:bestfit}. Figure~\ref{fig:energy_north_south} shows the deposited
energy spectra corresponding to the best-fit model parameters.
Figure~\ref{fig:zenith} shows the zenith angle distributions of the sample with
different energy thresholds.

\begin{table}[htbp]
	\caption{
	Best fit parameters and number of events attributable to each component.
The normalizations of the atmospheric fluxes are relative to the models
described in Sec.~\ref{sec:analysis_method}. The per-flavor normalization
$\Phi_0$ and spectral index $\gamma$ of the astrophysical flux are defined in
Eq.~\eqref{eq:powerlaw}; the fit to the astrophysical flux is sensitive to $25
\,\, \text{TeV} < E_{\nu} < 1.4 \,\, \text{PeV}$. The two-sided error ranges
given are 68\% confidence regions in the $\chi^2$ approximation; upper limits
are at 90\% confidence. The goodness-of-fit p-value for this model is $0.2$.
}
	\label{tab:bestfit}
	\centering
\begin{tabular}{lcc}
\hline
Parameter & Best-fit value & No. of events\\
\hline
Penetrating $\mu$ flux & $1.73 \pm 0.40 \, \Phi_{\textsc{sibyll}+\textsc{dpmjet}}$ & $30 \pm 7$ \\
Conventional $\nu$ flux & $0.97 ^{+0.10}_{-0.03} \, \Phi_{\textsc{HKKMS}}$ & $280^{+28}_{-8}$ \\
Prompt $\nu$ flux & $ < 1.52 \, \Phi_{\textsc{ERS}}$ (90\% CL) & $ < 23 $  \\
Astrophysical $\Phi_0$ & $2.06^{+0.35}_{-0.26} \times 10^{-18}$  & \multirow{3}{*}{$87^{+14}_{-10}$} \\
                       & ${\rm {GeV^{-1} \, cm^{-2} \, sr^{-1} \, s^{-1}} }$ \\
Astrophysical $\gamma$ & $2.46 \pm 0.12$ \\
\hline
\end{tabular}\end{table}

This simple model does not describe the data perfectly. There is a notable
departure in the southern sky around 30 TeV.
However, the excess is not statistically significant; correlated fluctuations
of the observed size or greater are expected from a smooth underlying power-law
spectrum in 5\% of experiments. The events in the energy and zenith region of the excess are
overwhelmingly cascade-like and display no signs of early hits from penetrating atmospheric
muons. Their rate far exceeds that expected from penetrating muon background
and conventional atmospheric neutrinos ($\sim 1$ event per year), and their
distribution in time and within the fiducial volume is compatible with a
uniform one. Known sources of systematic uncertainty in the neutrino acceptance of the
detector, like the optical properties of the glacial ice or the optical
efficiency of the DOM, are unable to create structure in the observed energy
distribution. At present, we interpret this as a statistical fluctuation.
We expect that future searches using more years of data will help constrain the
cause of the excess, either by reducing its significance or by
strengthening it enough that definitive statistical statements can be made.

\begin{figure}[htb]
	\centering
	
	\subfloat[][
		Likelihood profile in astrophysical power-law index $\gamma$ and
		per-flavor normalization $\Phi_0/10^{-18} \,\, \rm{GeV^{-1} \, cm^{-2} \, sr^{-1} \, s^{-1}}$.
		$E^{-2.5}$ is strongly favored over $E^{-2}$.
	]{
		\includegraphics[scale=1]{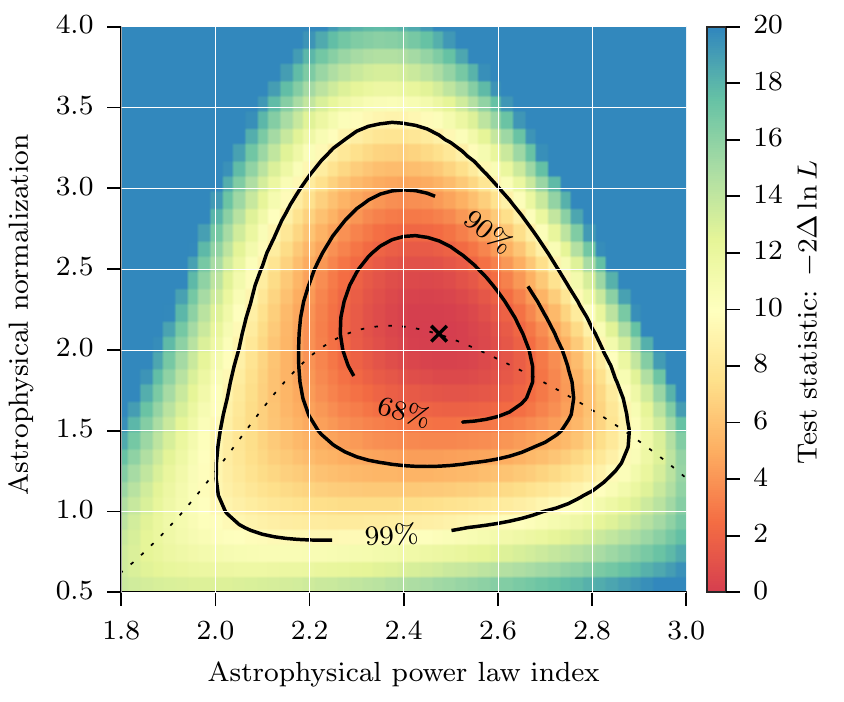}
		\label{fig:profile2d:index_astro}
	}
	
	\subfloat[][
		Likelihood profile in astrophysical power-law index $\gamma$ and prompt
		atmospheric neutrino flux normalization \cite{Sarcevic:2008}.
	]{
		\includegraphics[scale=1]{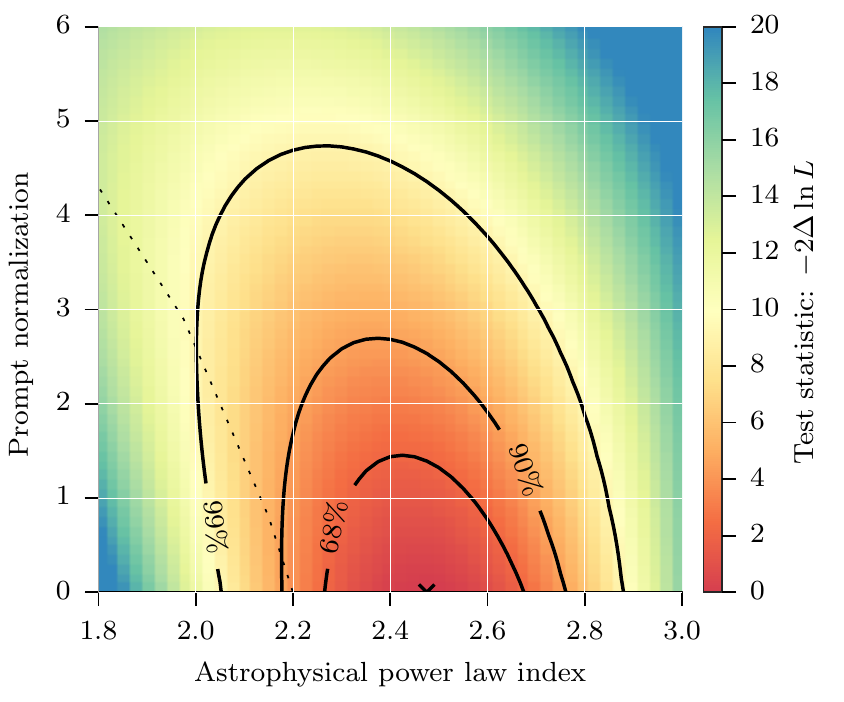}
		\label{fig:profile2d:index_prompt}
	}
		
	\caption{
	Profile likelihood scans showing the correlation between the
astrophysical power-law index and the normalizations of the astrophysical and
prompt atmospheric components. In each plot, the colors show the test statistic
\eqref{eq:test_statistic}, obtained by fixing the parameters shown on the axes
and varying all others to obtain the conditional best fit. The \textsc{x} shows the best-fit
point as in Tab.~\ref{tab:bestfit} and the contours show confidence regions in
the $\chi^2$ approximation \cite{wilks1938} with 2 degrees of freedom. The thin dotted line shows the
conditional best fit for each value of $\gamma$.
	}
	\label{fig:profile2d}
	
\end{figure}

The spectral index of $2.46$ needed to explain the low-energy data has
implications for the underlying neutrino production mechanism. As pointed out
in \cite{2013PhRvD..88l1301M}, $pp$ interactions produce neutrinos and
$\gamma$-rays that follow the same scale-free power-law spectrum, and the
$\gamma$ spectra from $pp$ interactions at $\sim$GeV energies can be
extrapolated to the TeV range where IceCube observes neutrinos. This
extrapolation argument does not apply to $p\gamma$ interactions. If the diffuse
extragalactic $\gamma$ background measured by Fermi-LAT is due to extragalactic
$pp$ interactions in optically thin regions, then the spectral index of the
associated neutrino spectrum must be
smaller than 2.2. \cite{2013PhRvD..88l1301M}. The data presented here indicate
that the neutrino spectrum is softer than $E^{-2.2}$ with 90\% confidence
(see Fig.~\ref{fig:profile2d:index_astro}), implying that one of these
assumptions is violated.

\begin{figure}[htb]
	\centering
		\includegraphics[scale=1]{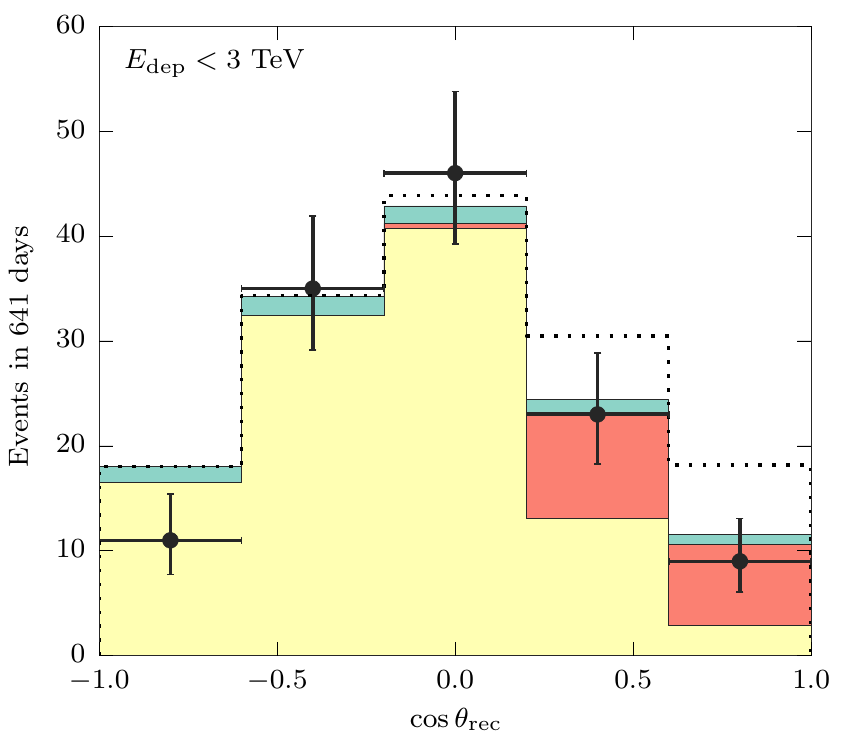}
	\caption{
	Verification of atmospheric neutrino veto with low-energy data. The points
show events depositing less than 3 TeV, while the stacked histograms show the
expected contributions from conventional atmospheric neutrinos, penetrating
muons, and the negligible contribution of astrophysical neutrinos, using the
color scheme of Fig.~\ref{fig:energy_north_south}. These match the observed
data much better than the the dotted line, which shows the number of events
that would be collected if atmospheric neutrinos were never vetoed by
accompanying muons.
	}
	\label{fig:zenith-3TeV}
\end{figure}

All of the parameters in Tab.~\ref{tab:bestfit} are correlated except for the
conventional atmospheric neutrino flux normalization. The latter
is determined mostly by the northern-sky data below 10~TeV deposited
energy, where the contributions of the other components are negligible, and is
compatible with the expected normalization \cite{Honda:2006} to within statistical errors,
providing a useful check of the neutrino acceptance calculated from simulation.
Similarly, the low-energy component provides a verification of the atmospheric
neutrino veto independent of the observed astrophysical excess, as shown in Fig.~\ref{fig:zenith-3TeV}.
The prompt atmospheric neutrino flux, on the other hand, can provide a significant
contribution to the overall event rate between 10 and 100 TeV deposited energy,
but has no region where it contributes exclusively. Its inferred
normalization depends on assumptions about the astrophysical neutrino flux. The
correlations between the astrophysical and prompt atmospheric components are
shown in Fig.~\ref{fig:profile2d}. Since the power-law index of the astrophysical flux is constrained
primarily by the large number of events below the pivot point at 100~TeV, the
normalization and index are correlated. Similarly, the prompt normalization is
correlated with the astrophysical index; as the index is forced to smaller
values, a larger prompt flux is required to explain the data between
10 and 100 TeV deposited energy. The normalization of the penetrating muon
component is constrained by the excess of southern-sky data over the conventional
atmospheric neutrino expectation below a few TeV, and is weakly correlated with
it (not shown).

\begin{figure}[htb]
	\centering
		\includegraphics[scale=1]{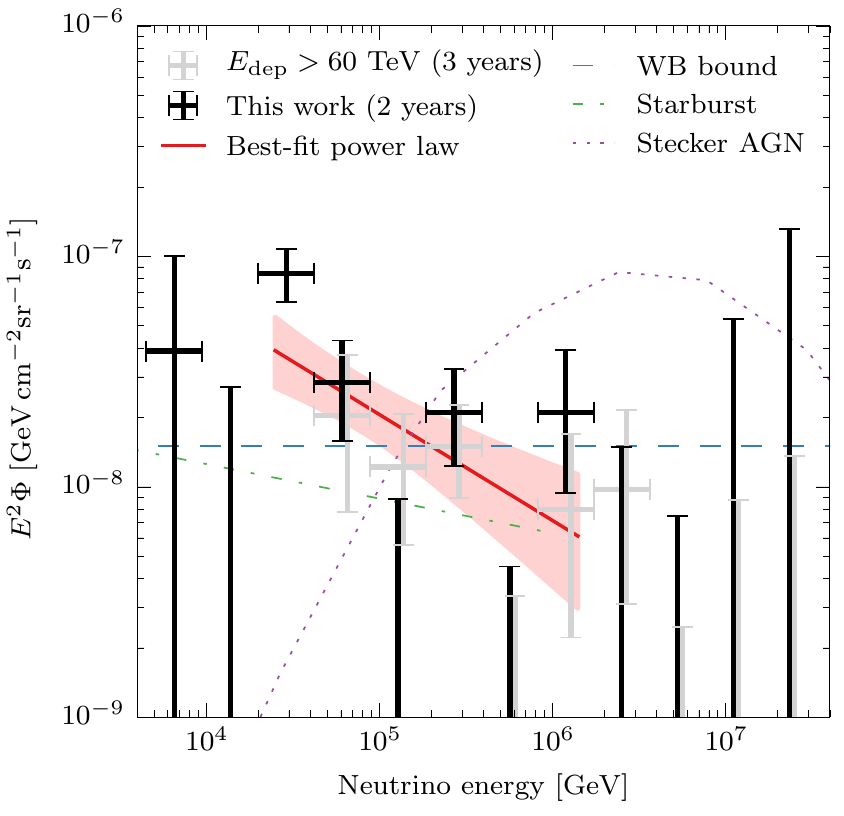}
	\caption{
	Unfolding the non-atmospheric excess as piecewise-constant per-flavor fluxes $E^2\Phi$.
	The horizontal error bars show the
range of primary neutrino energies that contribute to each bin, while the
vertical error bars show the range of $E^2\Phi$ that change the $-2\Delta
\ln L$ test statistic by less than 1. The black points show the fit to the data
sample presented here; the light grey data points are from the 3-year data
sample of \cite{HESE_3year}, shifted slightly to the right for better visibility.
Above the highest observed energy, the error bars
provide upper limits on the flux; these are less constraining than the upper limits
of \cite{EHE:2012} above 10 PeV.
The thin lines show models for the diffuse astrophysical
neutrino background: the upper bound from the total luminosity
of EeV cosmic rays from \cite{WaxmanBahcallBound} (blue),
the starburst galaxy model of \cite{WaxmanStarburst} (green),
and the AGN core emission model of \cite{SteckerErratum} (purple).
	}
	
	\label{fig:unfolding}
\end{figure}

These correlations would not be problematic if the model of the astrophysical
flux were exact, but since its sources are not known, any single model will
necessarily be an approximation. It is useful to examine how assumptions about
astrophysical models affect the upper limit on the prompt atmospheric
flux normalization. The first assumption made is that the astrophysical flux
must follow a single power-law energy distribution. This assumption can be
relaxed by describing the astrophysical neutrino flux with a piecewise constant
function of neutrino energy as shown in Fig.~\ref{fig:unfolding}. The observed
excess in the deposited energy spectrum is reflected in a corresponding excess
in the neutrino energy spectrum, and the additional freedom granted to the
astrophysical component weakens the 90\% upper limit on the prompt atmospheric
flux from 1.52 to 1.75 times the prediction of \cite{Sarcevic:2008}. This remaining limit is driven primarily by the
assumption that the astrophysical neutrino flux is isotropic. If this
assumption is weakened by allowing the astrophysical fluxes that contribute to
the northern- and southern-sky data to vary independently, the limit
relaxes further to 3.69. While this limit is not meaningfully smaller than the
previously published limit of 3.8 \cite{IC59Diffuse}, it involves many fewer
assumptions about the nature of the astrophysical neutrino background.

\section{Conclusion} 
\label{conclusion}

In the analysis presented here, we used a veto-based technique to isolate 388
events starting in the IceCube instrumented volume and depositing more than 1
TeV from 641 days of data, of which 92\% were neutrino events. Astrophysical
neutrino candidates were observed in the southern sky with energies as low as
10 TeV, far below the threshold of the previous high-energy starting event
analysis \cite{HESE, HESE_3year} and in a region inaccessible to the
traditional up-going track analysis \cite{IC59Diffuse}. We characterized the
contributions of penetrating atmospheric muons, conventional and prompt
atmospheric neutrinos, and astrophysical neutrinos to the data sample using a
likelihood fit to the distributions of deposited energy and zenith angle for
cascade and starting-track events.

The analysis yielded new information about the behavior of the neutrino
spectrum between 10 and 100 TeV. If the energy spectrum of the astrophysical
neutrinos is a single power law, then it must have a spectral index
of $2.46 \pm 0.12$, softer than the typical $E^{-2}$ benchmark spectrum. The $\gamma=2$
hypothesis can be rejected with 99\% confidence under this assumption.
The new constraint on the spectral index is due primarily to the lower deposited-energy
threshold of this analysis. If the deposited-energy threshold is raised to 60 TeV
(corresponding to sensitivity for $E_{\nu} > 100 \,\, \text{TeV}$),
then the best-fit spectral index hardens to $2.26 \pm 0.35$, compatible with the previous high-energy result \cite{HESE_3year}.
The statistically insignificant excess that appeared in the down-going data near 30 TeV, a region where atmospheric
leptons are heavily suppressed, had only a minor influence on the inferred spectral
index of the astrophysical neutrinos. If we force the spectral index to $\gamma=2$,
then the per-flavor normalization $\Phi_0$ (cf. Eq.~\eqref{eq:powerlaw}) drops to $1.22\pm0.5 \times 10^{-18} \,\, {\rm {GeV^{-1} \, cm^{-2} \, sr^{-1} \, s^{-1}} }$,
consistent with the previously published 90\% C.L. upper limit of 1.44 derived from northern-sky $\nu_{\mu}$ events \cite{IC59Diffuse}.
At the same time, we
searched for atmospheric neutrinos from charmed meson decay. No such component
was observed, and we placed upper limits on their flux. These limits depend
strongly on assumptions about the astrophysical neutrino background, and range from
1.52 times the prediction from perturbative QCD \cite{Sarcevic:2008} at 90\%
confidence when the astrophysical flux is assumed to follow a single
isotropic power-law distribution to 3.69 times the prediction when
it is described with a piecewise constant function of energy and zenith
angle.

The constraints on the astrophysical flux are currently limited by the small
number of observed high-energy events, making it difficult to draw strong
inferences about the classes of cosmic-ray accelerators from the characteristics
of the associated neutrino spectrum. Beyond astrophysical considerations, the
inability to model the astrophysical flux precisely and reliably extrapolate
its angular and energy distribution to lower energies impedes any attempt to
measure the level of charmed-meson production in air showers via high-energy
neutrinos. Both of these problems may be approached with more and different
data. IceCube will continue to collect data, and future iterations of this
analysis will be able to use at least twice as many high-energy neutrino events
to constrain the energy spectrum and eventually possible anisotropies of the
astrophysical neutrino flux. In addition to better modeling of the
astrophysical background, sensitivity to charmed meson production in the
atmosphere will be improved by analyzing penetrating muon events jointly with
neutrino events. These are produced in the same decays as prompt muon
neutrinos, but have no astrophysical background to contend with.

\begin{acknowledgements}
	We acknowledge the support from the following agencies:
	U.S. National Science Foundation-Office of Polar Programs,
	U.S. National Science Foundation-Physics Division,
	University of Wisconsin Alumni Research Foundation,
	the Grid Laboratory Of Wisconsin (GLOW) grid infrastructure at the University of Wisconsin - Madison, the Open Science Grid (OSG) grid infrastructure;
	U.S. Department of Energy, and National Energy Research Scientific Computing Center,
	the Louisiana Optical Network Initiative (LONI) grid computing resources;
	Natural Sciences and Engineering Research Council of Canada,
	WestGrid and Compute/Calcul Canada;
	Swedish Research Council,
	Swedish Polar Research Secretariat,
	Swedish National Infrastructure for Computing (SNIC),
	and Knut and Alice Wallenberg Foundation, Sweden;
	German Ministry for Education and Research (BMBF),
	Deutsche Forschungsgemeinschaft (DFG),
	Helmholtz Alliance for Astroparticle Physics (HAP),
	Research Department of Plasmas with Complex Interactions (Bochum), Germany;
	Fund for Scientific Research (FNRS-FWO),
	FWO Odysseus programme,
	Flanders Institute to encourage scientific and technological research in industry (IWT),
	Belgian Federal Science Policy Office (Belspo);
	University of Oxford, United Kingdom;
	Marsden Fund, New Zealand;
	Australian Research Council;
	Japan Society for Promotion of Science (JSPS);
	the Swiss National Science Foundation (SNSF), Switzerland;
	National Research Foundation of Korea (NRF);
	Danish National Research Foundation, Denmark (DNRF)
\end{acknowledgements}

\end{document}